\documentclass[conference]{IEEEtran}
\IEEEoverridecommandlockouts
\usepackage{cite}
\usepackage{amsmath,amssymb,amsfonts}
\usepackage{algorithmic}
\usepackage{graphicx}
\usepackage{textcomp}
\usepackage{xcolor}
\usepackage{booktabs}
\usepackage{hyperref}
\usepackage{url}
\def\BibTeX{{\rm B\kern-.05em{\sc i\kern-.025em b}\kern-.08em
    T\kern-.1667em\lower.7ex\hbox{E}\kern-.125emX}}
\begin{document}

\title{An Investigation Into Various Approaches For Bengali Long-Form Speech Transcription and Bengali Speaker Diarization}

\author{
\IEEEauthorblockN{Epshita Jahan}
\IEEEauthorblockA{\textit{Department of Computer Science and Engineering} \\
\textit{Bangladesh University of Engineering and Technology}\\
Dhaka, Bangladesh}
\and
\IEEEauthorblockN{Khandoker Md Tanjinul Islam}
\IEEEauthorblockA{\textit{Department of Computer Science and Engineering} \\
\textit{Bangladesh University of Engineering and Technology}\\
Dhaka, Bangladesh}
\and
\IEEEauthorblockN{Pritom Biswas}
\IEEEauthorblockA{\textit{Department of Computer Science and Engineering} \\
\textit{Bangladesh University of Engineering and Technology}\\
Dhaka, Bangladesh}
\and
\IEEEauthorblockN{Tafsir Al Nafin}
\IEEEauthorblockA{\textit{Department of Computer Science and Engineering} \\
\textit{Bangladesh University of Engineering and Technology}\\
Dhaka, Bangladesh}
}

\maketitle

\begin{abstract}
Bengali remains a low-resource language in speech technology, especially for complex tasks like long-form transcription and speaker diarization. This paper presents a multi-stage approach developed for the ``DL Sprint 4.0 | Bengali Long-Form Speech Recognition'' and ``DL Sprint 4.0 | Bengali Speaker Diarization'' \cite{b1} on Kaggle, addressing the challenges of ``who spoke when/what'' in hour-long recordings. We implemented Whisper Medium fine-tuned on Bengali data (bengaliAI/tugstugi\_bengaliai-asr\_whisper-medium) for transcription and integrated pyannote/speaker-diarization-community-1 with our custom-trained segmentation model to handle diverse and noisy acoustic environments. Using a two-pass method with hyperparameter tuning, we achieved a DER of 0.27 on the private leaderboard and 0.19 on the public one. For transcription, chunking, background noise cleaning, and algorithmic post-processing yielded a WER of 0.38 on the private leaderboard. These results show that targeted tuning and strategic data utilization can significantly improve AI inclusivity for South Asian languages.
All relevant codes are available at: \url{https://github.com/Short-Potatoes/Bengali-long-form-transcription-and-diarization.git}
\end{abstract}

\begin{IEEEkeywords}
Bengali speech recognition, speaker diarization, Whisper, ASR, low-resource languages, pyannote, voice activity detection
\end{IEEEkeywords}

\section{Introduction}
Recent advances in speech recognition have been driven by large-scale datasets and powerful models, yet low-resource languages like Bengali have not fully benefited. This study addresses that gap by proposing robust methodologies for long-form transcription and speaker diarization tailored to Bengali.

Long-form transcription is challenging because systems like Whisper impose a 30-second input limit, requiring segmentation of lengthy audio. To overcome this, we used voice activity detection (VAD) to identify speech regions, applied Whisper for transcription, and aligned outputs with post-processing to remove redundancies. Our dataset consisted of hour-long recordings with transcriptions, underscoring the need for effective strategies.

For speaker diarization, we fine-tuned Pyannote's segmentation model within its pipeline. A two-pass approach---coarse segmentation followed by refined analysis---improved accuracy. Evaluation was based on long-form audio with ground-truth speaker annotations in JSON format.

In summary, combining segmentation-based preprocessing, Whisper-based ASR, and fine-tuned diarization models significantly enhances speech recognition for Bengali and other low-resource languages.

\section{Methodology}

\subsection{Model Selection for Speaker Diarization}

Conversational Bangla diarization presents challenges including speaker overlap, acoustic variability, and unconstrained speaker counts. We evaluated three contemporary diarization frameworks: Pyannote Community \cite{b2}, NVIDIA NeMo Sortformer \cite{b3}, and Diarizen \cite{b4}.

Preliminary experiments included Pyannote Speaker Diarization v3.1; however, the newer \textit{community-1} pipeline demonstrates consistent improvements across multiple benchmarks. The framework integrates neural speaker embedding extraction with PLDA-based clustering \cite{b5}, providing robust speaker discrimination under domain variability. According to the official model benchmarks, reported diarization error rates (DER) include 11.7 on ALSHELL-4 and 11.2 on VoxConverse.

NeMo Sortformer employs an end-to-end transformer architecture but supports a maximum of four speakers per recording, constraining scalability in unconstrained conversational scenarios \cite{b3}. Diarizen reports performance gains over Pyannote v3.1; however, it remains less extensively validated in open-source benchmarking environments \cite{b4}.

Considering benchmark performance, clustering robustness, scalability, and implementation stability, we adopt the \textbf{Pyannote Community pipeline} as the diarization backbone.

\subsection{Model Selection For Automatic Speech Recognition (ASR)}

For Bangla ASR, we evaluated multiple transformer-based architectures built upon the Whisper and wav2vec2 frameworks.

Our baseline system employed bengaliAI/tugstugi\_bengaliai-asr\_whisper-medium, a Bangla fine-tuned Whisper-medium model, using manual chunking for long-form inference. This model was selected due to its strong performance on Bangla benchmarks and prior empirical validation in recent literature.

We further experimented with the WhisperX pipeline, combining: anuragshas/whisper-large-v2-bn, tugstugi\_bengaliai-asr\_whisper-medium, tanmoyio/wav2vec2-large-xlsr-bengali and ai4bharat/indicwav2vec\_v1\_bengali.

The WhisperX framework enables improved timestamp alignment and segmentation through VAD-based chunking and alignment refinement. However, \textbf{tugstugi\_bengaliai-asr\_whisper-medium} consistently yielded the lowest word error rate (WER). Accordingly, we adopted this model as our primary ASR backbone \cite{b6}.

\subsection{Preprocessing for Speaker Diarization}

All audio recordings were preprocessed using \textbf{DEMUCS} for speech enhancement. The model was applied to suppress background noise and isolate speech components prior to diarization.

No additional spectral transformations or aggressive filtering were applied in order to preserve speaker-discriminative acoustic characteristics.

\subsection{Preprocessing for ASR}

We evaluated both speech enhancement and segmentation strategies prior to transcription. Although DEMUCS-based denoising was initially explored, it did not improve performance and was therefore excluded from the final system.

For segmentation, we compared Silero VAD with a fine-tuned task-specific segmentation model. The fine-tuned model provided more accurate boundary detection and was adopted in the final ASR pipeline. No additional feature engineering was applied, as the transformer-based model operates directly on waveform-derived log-Mel features.

\subsection{Training for Diarization}

We trained only the segmentation component of the Pyannote diarization-3.1 model for this task. Several training strategies were explored. Initially, we trained the segmentation model exclusively on the competition dataset, which yielded our best performance. We also experimented with a larger multilingual dataset that combined the competition's Bengali corpus with an additional Hindi dataset. Our rationale was that speaker diarization is generally considered language-agnostic, and thus, the inclusion of more diverse data might improve generalization. However, this approach resulted in slightly worse performance compared to training solely on the Bengali dataset.

The training experiments varied primarily in three dimensions: learning rate, batch size, and regularization strength. Lower learning rates with moderate batch sizes and minimal weight decay produced the most stable convergence and ultimately the best diarization scores. Increasing the learning rate and regularization, while reducing batch size, accelerated training but introduced instability and degraded performance. Extending the number of epochs provided marginal improvements but did not surpass the stability achieved with conservative hyperparameters. Overall, the most effective configuration balanced gradual learning with limited regularization, highlighting the sensitivity of diarization segmentation to training dynamics.

In addition to training-time optimization, we performed a structured grid search over clustering and post-processing hyperparameters using the training set as a development set. The search was conducted in three phases: (i) clustering threshold sweep with single-pass inference, (ii) caching raw predictions at the optimal threshold, and (iii) post-processing parameters sweep (e.g., minimum duration and merge gaps) without re-running the full pipeline. This staged strategy enabled efficient hyperparameter tuning while minimizing redundant inference overhead.

\subsection{Training Long Form Transcription}

For long-form transcription, we attempted to fine-tune the ASR model using an external dataset. This approach, however, yielded extremely poor results, most likely due to inconsistencies and noise within the training data. To address this, we integrated the segmentation model previously trained for diarization into our ASR pipeline. The segmentation model was used to generate coherent audio chunks, which were then transcribed by the ASR system. This strategy provided a more reliable framework for handling extended recordings, as the diarization-informed segmentation improved the alignment and consistency of transcription outputs.

\begin{figure}[htbp]
\centering
\includegraphics[width=0.45\textwidth]{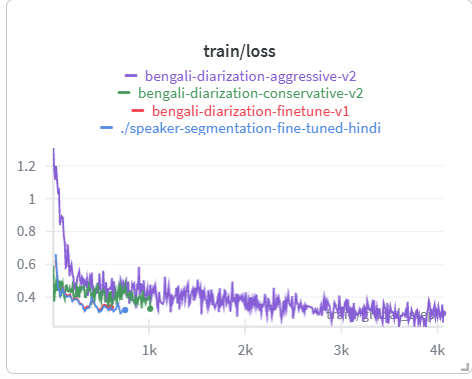}
\includegraphics[width=0.45\textwidth]{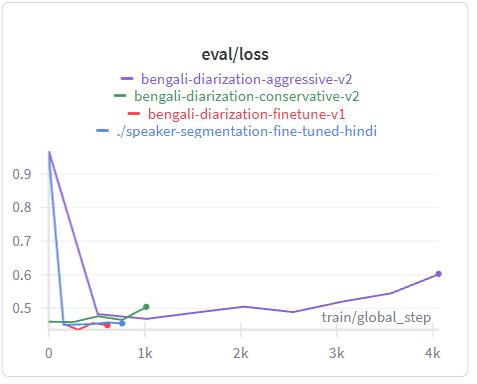}
\caption{Diarization performance results showing training progression and comparative analysis.}
\label{fig:diarization_results}
\end{figure}

\section{Results}

\subsection{Results For Diarization}

After training for approximately 3--4 epochs, our segmentation model achieved a substantial improvement over the baseline diarization error rate (DER) of 0.40, reducing it to 0.199 on the public evaluation set. Building upon this, we applied a two-pass decoding strategy combined with post-processing refinements, which further lowered the public set DER to 0.192. On the private evaluation set, the same approach yielded a DER of 0.27. These results demonstrate that iterative training and multi-stage inference can significantly enhance diarization performance, particularly on the public benchmark, while still maintaining competitive results on the private dataset.

\subsection{Results for Automatic Speech Recognition}

Using the \textit{Tugstugi Whisper-medium} model with \textit{Silero VAD} segmentation \cite{b7}, our initial system achieved a WER of \textbf{0.36771} on the public evaluation set. Incorporating \textit{Qwen LLM}-based post-processing increased the WER to \textbf{0.37921}, while adding \textit{DEMUCS}-based speech enhancement resulted in a WER of \textbf{0.37614}, indicating no improvement over the baseline configuration. Finally, replacing Silero VAD with our \textit{fine-tuned segmentation} model yielded the best public performance, reducing the WER to \textbf{0.36674}.

On the private evaluation set, the corresponding configurations produced WERs of \textbf{0.38970} (\textit{Tugstugi + Silero}), \textbf{0.39657} (adding \textit{Qwen LLM}), \textbf{0.39426} (adding \textit{DEMUCS}), and \textbf{0.38005} (\textbf{fine-tuned segmentation}), with the fine-tuned segmentation model again achieving the strongest result.

\begin{table*}[t]
\caption{Diarization Methods and Error Rates (DER)}
\begin{center}
\begin{tabular}{|l|c|}
\hline
\textbf{Method} & \textbf{Error Rate (DER for Private)} \\
\hline
\textbf{Pyannote} (Finetuned \texttt{lucius-40/speaker-segmentation-fine-tuned-bn}) + Demucs (using two-pass) & 0.28286 \\
\hline
\textbf{Pyannote} (Finetuned \texttt{lucius-40/speaker-segmentation-fine-tuned-bn}) + Demucs & 0.27318 \\
\hline
\textbf{Pyannote} (Finetuned \texttt{lucius-40/speaker-segmentation-fine-tuned-bn}) & 0.27773 \\
\hline
\textbf{Pyannote} (FineTuned: \texttt{speaker-segmentation-bengali-optimized-cfg-1-balanced}) & 0.27346 \\
\hline
\textbf{Pyannote} (FineTuned: \texttt{speaker-segmentation-fine-tuned-hindi-and-english}) & 0.28413 \\
\hline
\textbf{Pyannote} (Finetuned \texttt{lucius-40/speaker-segmentation-bengali-optimized-conservative}) & 0.25921 \\
\hline
\textbf{Pyannote} (Finetuned \texttt{lucius-40/speaker-segmentation-fine-tuned-bn}) with hypertuned parameters (using two-pass) & 0.26388 \\
\hline
\textbf{Pyannote} (Finetuned \texttt{lucius-40/speaker-segmentation-bengali-optimized-conservative}) + Demucs (using two-pass) & 0.27988 \\
\hline
DiariZen & 0.46624 \\
\hline
\end{tabular}
\label{tab:diarization_results}
\end{center}
\end{table*}

\begin{table*}[t]
\caption{Long-form Speech Detection Methods and Error Rates (WER)}
\begin{center}
\begin{tabular}{|p{12cm}|c|}
\hline
\textbf{Method} & \textbf{Error Rate (WER)} \\
\hline
\texttt{bengaliAI/tugstugi\_bengaliai-asr\_whisper-medium} + \texttt{lucius-40/speaker-segmentation-fine-tuned-bn} & 0.38005 \\
\hline
\texttt{bengaliAI/tugstugi\_bengaliai-asr\_whisper-medium} + Silero VAD & 0.38970 \\
\hline
Demucs + \texttt{lucius-40/speaker-segmentation-fine-tuned-bn} + \texttt{bengaliAI/tugstugi\_bengaliai-asr\_whisper-medium} & 0.39426 \\
\hline
\texttt{bengaliAI/tugstugi\_bengaliai-asr\_whisper-medium} + Silero VAD + Qwen2.5-7B Few-Shot Correction & 0.39657 \\
\hline
Whisper-X & 0.79 \\
\hline
\end{tabular}
\label{tab:asr_results}
\end{center}
\end{table*}

\section{Discussion}

\subsection{Effect of using WhisperX with Bengali fine-tuned Whisper and alignment model}

We tried to use the WhisperX pipeline along with \textit{tugstugi\_bengaliai-asr\_whisper-medium} as the base model and \textit{tanmoyio/wav2vec2-large-xlsr-bengali} and \textit{ai4bharat/indicwav2vec\_v1\_bengali} as alignment models. They yielded very poor results, with a baseline WER of 0.79. The alignment model did not work as intended to produce clean transcriptions.

\subsection{Effect of using Demucs to filter background noise}

We used the Demucs library to filter out background noise for both tasks. It yielded a noticeable improvement in the diarization task, clearly helping to detect speaker segments more effectively. However, using it in long-form ASR did not yield any noticeable positive results, possibly due to mismatch between enhanced signals and Whisper's pretraining distribution \cite{b8}.

\subsection{Effect of task-specific segmentation}

Task specific segmentation is more beneficial than generic VAD. The fine-tuned segmentation model achieved the lowest WER and DER respectively in each task.

\subsection{Effect of the two-pass approach in diarization}

For diarization, we used our pipeline to detect the number of speakers in the given audio file. We then passed the number of speakers as a fixed constraint on the next pass, which helped reduce hallucinations. This approach yielded extremely good results and improved the DER from an initial 0.199 to 0.192. This improvement shows that explicitly passing the number of speakers to the pipeline before inference significantly enhances accuracy.

\subsection{Effect of Post-Processing}

Post-processing had a positive impact on both the diarization and transcription tasks. For the transcription task, we noticed that the raw ASR output tended to repeat words and phrases frequently, most likely due to confusion with background music. To address this, we used an algorithm that removes repetitions at the word, phrase, and letter levels (while preserving natural repetitions). For the diarization task, we removed A-B-A segments and merged them into A-A if B had a very short duration, as it could be a hallucination by the model. Similarly, we removed very small segments and merged same-speaker segments separated by very short silences. These steps yielded noticeable improvements.

\section{Limitations and Ethical Considerations}

Although performance improvements were achieved, several limitations remain. In the diarization pipeline, only the segmentation component was fine-tuned, while the speaker embedding model was not trained on task-specific data. Attempts to fine-tune a WeSpeaker embedding model did not yield improvements, and substituting ECAPA-TDNN embeddings resulted in degraded performance, likely due to padding and alignment inconsistencies. Additionally, alternative clustering strategies were not systematically explored due to time constraints.

For ASR, although task-specific fine-tuning is generally known to improve performance, our fine-tuning attempts resulted in degraded accuracy compared to the non-fine-tuned model, likely due to pipeline and methodological inconsistencies.

From an ethical perspective, diarization systems may be extended toward speaker profiling applications, raising potential privacy concerns. Furthermore, the system may exhibit inherent biases arising from dataset imbalance, which could affect performance across different speaker demographics or dialects.

\section{Conclusion}

We present an effective framework for Bangla speaker diarization and automatic speech recognition by leveraging transformer-based architectures and task-specific segmentation strategies.

Previous diarization approaches have largely relied on conventional clustering pipelines and fixed segmentation modules. Our work demonstrates the effectiveness of modern neural diarization frameworks combined with fine-tuned segmentation for improved boundary detection in conversational Bangla speech. Although the Pyannote framework provides a robust end-to-end pipeline, we were unable to separately fine-tune the speaker embedding model due to computational and time constraints. We believe that domain-adaptive training of the embedding extractor could further reduce diarization error rates.

For ASR, we adopt a Whisper-based architecture fine-tuned for Bangla. While segmentation improvements yielded measurable gains, the acoustic model itself was not further fine-tuned on the competition dataset. Incorporating task-specific fine-tuning with CTC-based alignment or hybrid alignment strategies could potentially enhance phoneme-level consistency and reduce word error rates. We anticipate that such alignment-aware optimization would further strengthen transcription robustness in spontaneous conversational settings.

\section{Contributions}

Our work introduces a practical multi-stage framework for Bangla diarization and ASR. For diarization, we implement a two-pass inference strategy, where Pyannote is first run to estimate the maximum number of speakers, followed by a second pass with the inferred speaker count constrained in the pipeline, improving speaker assignment stability. We further enhance performance by fine-tuning the segmentation component on task-specific data.

For ASR, we incorporate segmentation-aware decoding and apply n-gram repetition removal as a lightweight post-processing step to reduce redundant token generation. Together, these modifications provide consistent improvements without altering the core pretrained backbones.

\end{document}